\documentclass[fleqn,twoside]{article}
\usepackage{espcrc2}

\usepackage{graphicx}
\usepackage[figuresright]{rotating}

\newcommand{\AmS}{{\protect\the\textfont2
  A\kern-.1667em\lower.5ex\hbox{M}\kern-.125emS}}

\title{The Chiral Limit of Non-compact QED$_3$}

\author{S.J. Hands \address{Department of Physics,
University of Wales Swansea, Singleton Park, Swansea, SA2 8PP, UK},
        J.B. Kogut \address{Department of Physics, 
University of Illinois at Urbana-Champaign, Urbana, IL 61801-3080, USA}, 
L. Scorzato \address{Deutsches Elektronen-Synchrotron DESY, Notkestr. 85, 
D-22603 Hamburg, Germany} and
C.G. Strouthos $^{\rm a}$ 
}

\begin{document}

\begin{abstract}
Non-compact QED$_3$ with four-component 
fermion flavor content $N_f \geq 2$ is studied numerically near
the chiral limit 
to understand its chiral symmetry breaking features. 
We monitor discretization and finite size effects on the chiral condensate 
by simulating the model at different values of the gauge coupling on lattices 
ranging in size 
from $10^3$ to $50^3$. Our upper bound for the dimensionless condensate 
$\beta^2\langle\bar\Psi\Psi\rangle$ in 
the $N_f=2$ case is $5 \times 10^{-5}$.
\vspace{1pc}
\end{abstract}

\maketitle

\section{Introduction}
Over the last few years, QED$_3$ has attracted a lot of attention
because of its potential applications to models of high $T_c$
superconductivity. It is believed to be confining and exhibit features such as dynamical mass
generation when the number of fermion flavors $N_f$ is smaller than a critical
value $N_{fc}$.
It is therefore an interesting and  challenging model 
and an ideal laboratory in which to study more complicated gauge field theories.

We are considering the four-component formulation of QED$_3$ where the Dirac 
algebra is represented by the $4 \times 4$ matrices $\gamma_0$, $\gamma_1$ 
and $\gamma_2$. This formulation preserves parity and gives each spinor 
a global $U(2)$ symmetry generated by
$\bf1$, $\gamma_3, \gamma_5$ and $i\gamma_3 \gamma_5$; the full 
symmetry is then $U(2N_f)$. If the fermions acquire dynamical mass 
the $U(2N_f)$ symmetry is broken spontaneously to $U(N_f) \times U(N_f)$
and $2N_f^2$ Goldstone bosons appear in the particle spectrum.  

At the present time the issue of spontaneous chiral 
symmetry breaking in QED$_3$ is not very well understood.
Studies based on Schwinger-Dyson equations (SDEs) using the
photon propagator derived from the leading order $1/N_f$ expansion
suggested that for $N_f$ less than some critical value $N_{fc}$
the answer is positive
with $N_{fc}\simeq3.2$ \cite{pisarski}.
Apparently, for $N_f > N_{fc}$, the attactive interaction between a fermion and 
an antifermion due to photon exchange is overwhelmed by the fermion 
screening of the theory's electric charge.
More recent studies which treat
the vertex consistently in both numerator and denominator of the SDEs 
have found a value for $N_{fc}$ either in agreement with the
original study
or slightly higher, with $N_{fc}\simeq4.3$ \cite{maris}.
Finally an argument based on a thermodynamic inequality has predicted
$N_{fc}\leq{3\over2}$ \cite{appelquist}.
There have also been numerical attempts to resolve the issue 
via lattice simulations. The obvious advantage in this 
case is that one can study any $N_f$ without any assumption 
concerning the convergence of expansion methods.
However, the principal obstruction to a definitive answer has been 
large finite volume effects resulting from the presence of a 
massless photon. 
Numerical studies of the quenched case have shown that chiral symmetry 
is broken \cite{quenched}, whereas in the case of simulations
with dynamical fermions opinions have divided on whether $N_{fc}$ is finite and $\approx 3$
or whether chiral symmetry is broken for all $N_f$ \cite{dynamical}. 
Recent studies of the $N_f=1$ model on small lattices appear
in \cite{maris2002}.

In our study we used the Hybrid Monte Carlo algorithm to simulate 
the non-compact version of lattice 
QED$_3$ with staggered fermions. In the continuum it corresponds 
to the four-component 
spinor formulation of the model \cite{burden}. 
We implement even-odd partitioning which implies that a single flavor of 
one-component staggered 
fermions can be simulated, which corresponds to  
$N_f=2$ in the continuum limit.  

\section{Results}
In this section we discuss the results of lattice simulations
of QED$_3$ with $N_f \geq 2$. In our study we tried to detect
and control the various drawbacks of the lattice method:
(i) The lattice itself distorts continuum space-time physics
considerably unless the lattice spacing $a$ can be chosen small compared
to the relevant physical wavelengths in the system.
(ii) the size of the lattice $L^3$ must be large compared to the
dynamically generated correlations in the system; and
(iii) the chiral limit can only be studied
by simulating light fermions of mass $m_0$ in lattice units.

\begin{figure}[htb]
\centering
\includegraphics[scale=0.57]{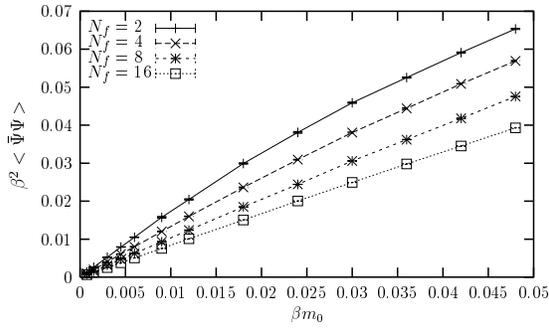}
\vspace{-1cm}
\caption{Dimensionless condensate $\beta^2 \langle \bar\Psi\Psi \rangle$
vs. dimensionless bare mass $\beta m_0$ for $N_f=2,4,8,16$, $\beta=0.6$
on a $16^3$ lattice.}
\label{fig:all_Nf}
\end{figure}
\vspace{-0.5cm}

In Fig.\ref{fig:all_Nf} we plot the
dimensionless chiral condensate $\beta^2 \langle \bar\Psi\Psi \rangle$ vs. the dimensionless bare mass
$\beta m_0$ (where $\beta\equiv \frac{1}{g^2a}$) for $N_f=2,4,8,16$.
The coupling $\beta=0.6$ and the lattice volume is $16^3$.
As $N_f$ increases the chiral condensate decreases (for $m_0 \geq 0$) because
the interaction between the fermion and the
antifermion is screened. However,
as $m_0 \rightarrow 0$
all the curves tend to pass smoothly through the origin.
This motivated us to study in more detail the pattern of chiral symmetry
breaking at small $N_f$ on larger volumes near the chiral limit.
\begin{figure}[htb]
\centering
\includegraphics[scale=0.57]{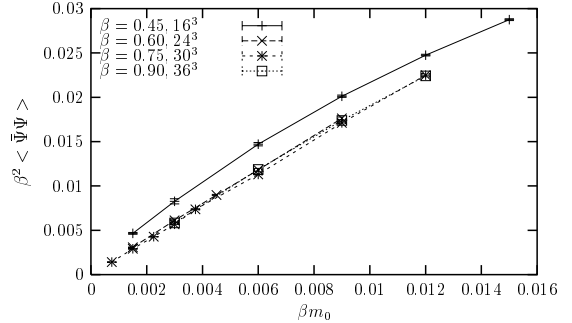}
\vspace{-1cm}
\caption{Dimensionless condensate vs. dimensionless bare mass for $N_f=2$ at different
values of the coupling $\beta$ and constant physical volume $(L/\beta)^3$.}
\label{fig:discret}
\end{figure}

In order to check whether our lattice data are characteristic of the continuum limit we
plot
in Fig.\ref{fig:discret} $\beta^2 \langle \bar\Psi\Psi \rangle$
vs. $\beta m_0$ for $N_f=2$ and coupling
$\beta=0.45,0.60,0.75,0.90$.
To disentangle the lattice discretization effects from the finite size effects we keep
the volume in physical units $(L/\beta)^3$ constant.
It can be inferred from the graph that 
discretization  effects are small for $\beta \geq 0.60$, 
because the data almost fall on the same line within
the resolution of our analysis.

\begin{figure}[htb]
\centering
\includegraphics[scale=0.57]{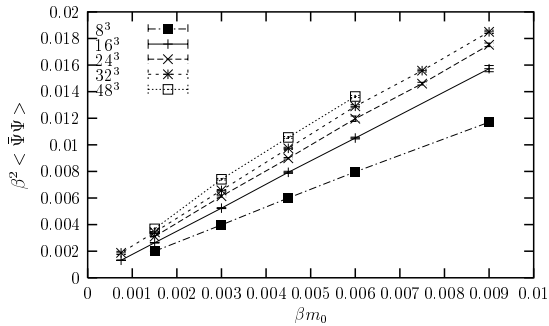}
\vspace{-1cm}
\caption{Dimensionless condensate vs. dimensionless bare mass for $N_f=2$, $\beta=0.6$
and lattice sizes $8^3, 16^3, 24^3, 32^3, 48^3$.}
\label{fig:cond_b=0.6}
\end{figure}

In Fig.\ref{fig:cond_b=0.6} we present our results for the chiral condensate
vs. bare mass for $N_f=2$ and $\beta=0.6$ on lattice sizes varying from $8^3$
to $48^3$.
We infer that finite size effects
become small for $L \geq 24$ and all the lines tend to pass smoothly through
the origin. Our analysis of meson masses and susceptibilities in scalar and 
pseudoscalar 
channels showed that 
these quantities suffer from very strong finite size effects
and therefore did not allow us to reach such
a definitive conclusion \cite{qed3.2002}.

\begin{figure}[htb]
\centering
\includegraphics[scale=0.57]{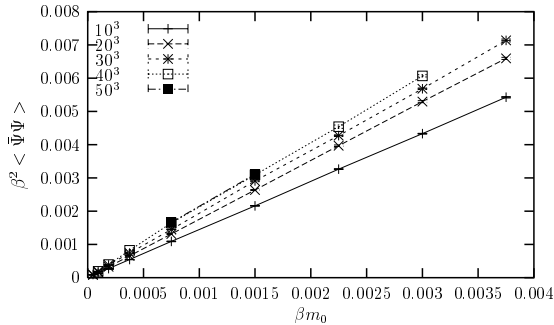}
\vspace{-1cm}
\caption{Condensate vs. bare mass for $N_f=2$, $\beta=0.75$ and lattice sizes $10^3,
20^3, 30^3, 40^3$ and $50^3$.}
\label{fig:cond_b=0.75}
\end{figure}

Next we discuss the results from $N_f=2$ simulations at $\beta=0.75$.
The $\beta=0.75$ data set is closer to
the continuum limit than the data extracted at $\beta=0.60$.
However, particular care is required because weak coupling data are very sensitive to finite
size effects and accurate measurements require simulations on large lattices.
In Fig.\ref{fig:cond_b=0.75} we present the results for the chiral
condensate vs. fermion bare mass
extracted from simulations with lattice sizes ranging from $10^3$ to $50^3$.
These simulations were performed very close to the chiral limit, i.e.
with $m_0 \leq 0.005$.
We can see from the figure the finite size effects are under
relatively good control and the data tend to pass smoothly through the origin.
Therefore, we conclude that for $N_f=2$ $\beta^2 \langle \bar \Psi \Psi \rangle \leq 5 \times 10^{-5}$,
which is a strong indication
that QED$_3$ may be chirally symmetric for $N_f \geq 2$.
\vspace{-0.5cm}
\vspace{-0.5cm}

\section{Conclusions}
In our study of QED$_3$ with $N_f \geq 2$
we attempted to establish whether chiral symmetry is broken or not
by studying 
the behavior of the chiral condensate
close to the continuum limit
$g \rightarrow 0$, on different  volumes in order to detect and control
finite size effects and near the chiral limit
$m_0 \rightarrow 0$. 

Our upper bound for the condensate in the $N_f=2$ case is
$\beta^2\langle \bar \Psi \Psi \rangle \leq 5 \times 10^{-5}$
and all the lines of $\beta^2 \langle \bar \Psi \Psi \rangle$ vs.
$\beta m_0$ tend to pass smoothly through the origin which may
imply that chiral symmetry is restored for $N_f \geq 2$.
We are continuing this study to check if chiral symmetry is broken 
in the case of $N_f=1$.

\section*{Acknowledgements}
SJH and CGS were supported by the Leverhulme Trust. 
JBK was supported in part by NSF grant PHY-0102409. 
The computer simulations were done on the Cray SV1's at NERSC, the IBM-SP
at NPACI,
and on the SGI Origin 2000 at the University of Wales Swansea.

\end{document}